\documentclass[sort&compress,final,5p,twocolumn,times]{elsarticle}

\journal{Physics Letters B}

\usepackage{graphicx,bm,color}
\usepackage[dvipsnames]{xcolor}
\usepackage{amsmath}
\usepackage{amssymb}
\usepackage{booktabs}
\usepackage{float}
\usepackage[caption=false]{subfig} 

\usepackage[hidelinks]{hyperref}
 \hypersetup{
   pdftitle={},%
   pdfauthor={},%
   pdfsubject={},%
   pdfkeywords={},%
   pdfstartview={},%
   bookmarksopen=true, breaklinks=true, debug=true, %
   colorlinks=true, linkcolor=red, citecolor=blue, urlcolor=blue
 }
\usepackage[normalem]{ulem}

\newcommand{\pjT}{p_{{\text j}T}}

\newcommand*\oline[1]{%
   \vbox{%
     \hrule height 0.5pt
     \kern0.4ex
     \hbox{%
       \kern-0.15em
       \ifmmode#1\else\ensuremath{#1}\fi
       \kern-0.15em
     }
   }
}

\begin{document}

\title{Collins function for pion-in-jet production in polarized $pp$ collisions: \\ a test of universality and factorization}

\date{\today}

\author[add1,add2]{Umberto D'Alesio}
\ead{umberto.dalesio@ca.infn.it}

\author[add1,add2]{Carlo Flore\corref{cor1}}
\ead{carlo.flore@unica.it}
\cortext[cor1]{Corresponding author}

\author[add3]{Marco Zaccheddu}
\ead{zacch@jlab.org}

\address[add1]{Dipartimento di Fisica, Universit\`a di Cagliari, Cittadella Universitaria, I-09042 Monserrato (CA), Italy}
\address[add2]{INFN, Sezione di Cagliari, Cittadella Universitaria, I-09042 Monserrato (CA), Italy}
\address[add3]{Theory Center, Jefferson Lab, 12000 Jefferson Avenue, Newport News, Virginia 23606, USA}

\begin{abstract}
We present an updated study of the Collins azimuthal asymmetries for pion-in-jet production in polarized $pp$ collisions. To this aim, we employ a recent extraction of the transversity and Collins fragmentation functions from semi-inclusive deep inelastic scattering and $e^+ e^-$ annihilation into hadron pairs processes, obtained within a simplified transverse momentum dependent (TMD) approach at leading order in  the strong coupling constant $\alpha_s$. In the present case we adopt a collinear configuration for the initial state, keeping transverse momentum effects only in the fragmentation mechanism. Our theoretical estimates, when compared against 200~GeV and 510~GeV data from the STAR Collaboration, show a generally good agreement for the distributions in the transverse momentum of the jet, the pion longitudinal momentum fraction and its transverse momentum with respect to the jet direction. While not being a proof, due the assumptions and limitations behind the entire approach, these findings corroborate the hypothesis of TMD factorization for such processes as well as of the universality of the Collins function and, once again, of a reduced impact of the proper TMD evolution on azimuthal asymmetries.

We will also present predictions based on an extraction of the Collins and transversity distributions where information from data on single spin asymmetry for inclusive pion production in $p^\uparrow p$ collisions is included through a Bayesian reweighting procedure.
\end{abstract}

\begin{keyword}
Collins and transversity functions \sep Azimuthal Asymmetries \sep Universality \sep JLAB-THY-25-4388
\end{keyword}

\maketitle
\section{\label{sec:intro}Introduction}

Understanding the internal structure of high-energy nucleons in terms of parton degrees of freedom, as well as the fragmentation mechanism of partons into hadrons, are certainly an utmost issue in hadron physics and more generally in QCD. When also intrinsic transverse momentum and spin effects are included the richness of information one can extract becomes invaluable. In this respect, transverse single-spin and azimuthal asymmetries in inclusive and semi-inclusive processes represent a unique tool to investigate these issues.

A well defined and consolidated framework to deal with these observables is the so-called transverse momentum dependent (TMD) approach~\cite{Kotzinian:1994dv,Tangerman:1994eh,Tangerman:1995hw,Boer:1997nt,Boer:1997mf,Bacchetta:2006tn,Arnold:2008kf,Pitonyak:2013dsu}, whose factorization allows one to separate the soft physics from the hard, perturbatively calculable, parts. There is a general consensus on the validity of TMD factorization for some specific processes, namely semi-inclusive deep inelastic scattering (SIDIS), Drell-Yan (DY) processes and $e^+e^-$ annihilation into almost back-to-back hadron pairs, where two ordered energy scales are present~\cite{Collins:1981uk,Collins:1984kg, Ji:2004wu, Collins:2011zzd,Echevarria:2012js}. These are the large virtuality of the exchanged boson, and a small one, identified with the transverse momentum of the final hadron in SIDIS, of the lepton pair in DY or the imbalance in the transverse momentum of the hadron pair in $e^+e^-$ processes.

Thanks to the significant experimental activity carried out during the last two decades~\cite{Belle:2008fdv,COMPASS:2008isr,HERMES:2009lmz,JeffersonLabHallA:2011ayy,COMPASS:2012dmt,BaBar:2013jdt}, this theoretical approach has been successfully adopted in several phenomenological analyses. Moreover, it is currently and will be further investigated in ongoing and future experimental programs at  JLab~\cite{Dudek:2012vr, Accardi:2023chb}, RHIC~\cite{Aschenauer:2015eha} at BNL, COMPASS/AMBER~\cite{COMPASS:2010shj,Bradamante:2018ick,Adams:2018pwt} at CERN, BABAR~\cite{BaBar:2013jdt} at SLAC, Belle II~\cite{Belle:2008fdv} at KEK and BESIII~\cite{BESIII:2015fyw}, in fixed-target mode at the LHC~\cite{Aidala:2019pit,LHCspin:2025lvj}, at the Electron-Ion Collider (EIC)~\cite{Boer:2011fh,Accardi:2012qut} and at Tevatron with the SpinQuest~\cite{SeaQuest:2019hsx} Drell-Yan program. 

These combined efforts from theory, experiments and phenomenology, have eventually allowed for the extraction, with an increasing level of accuracy, of important TMDs: among them we mention the Sivers distribution~\cite{Sivers:1989cc,Sivers:1990fh} and the Collins fragmentation function (FF)~\cite{Collins:1992kk}. The latter TMD, on which we concentrate in this work, describes the asymmetric azimuthal distribution of an unpolarized hadron around the direction of motion of the transversely polarized fragmenting parent quark. Moreover, it is chiral odd and naively T-odd. The Sivers and the Collins functions share important features, but while the first one is expected to be process dependent, even if in a predictable way when moving from SIDIS to DY processes~\cite{Collins:2002kn,Brodsky:2002rv,Boer:2003cm}, the second one is predicted to be universal~\cite{Metz:2002iz,Boer:2003cm,Collins:2004nx,Yuan:2009dw,Boer:2010ya}. 

In this paper we will focus on this fundamental aspect of the Collins function, studying a  process complementary to the three TMD factorizable ones mentioned above: the azimuthal asymmetry for pion-in-jet production in polarized $pp$ collisions. This process shares indeed a common feature with the other ones, the presence of two ordered energy scales: a large one, the transverse momentum of the jet, $p_{{\rm j}T}$, and a small one, the intrinsic transverse momentum of the pion with respect to the jet direction ($p_{\perp\pi}\, \equiv j_T$ in the following). In this regard, the hypothesis of TMD factorization is expected to be a reliable assumption.  

This study will eventually allow us to test this issue, or better said, to investigate the absence of sizable factorization breaking effects, as well as the role and the importance of the TMD evolution.
On this basis we use the information on the Collins function and the transversity distribution, as extracted from azimuthal asymmetries in SIDIS and $e^+e^-$ processes, to give theoretical estimates for the Collins asymmetry for pion-in-jet production in $p^\uparrow p$ collisions and compare them with available experimental data.

This work is an extended and improved study of analogous analyses carried out in the past~\cite{Yuan:2007nd,DAlesio:2010sag,DAlesio:2017bvu,Kang:2017btw,Kang:2017glf}, where similar conclusions were obtained, even if adopting less accurate TMD extractions and considering a limited number of data and with lower statistics, as available at that time. Here, the full set of data by the STAR Collaboration at $\sqrt s =$ 200 GeV~\cite{STAR:2022hqg} and the latest ones at $\sqrt s = 510$ GeV~\cite{Zhang:2024zuq,STAR:2025xyp}, suitably binned and providing $j_T$ distributions, and adopting updated TMD extractions, will allow us to perform a more detailed analysis. In this respect, even if on a phenomenological basis and taking into account the assumptions adopted in our simplified scheme, this work represents an important step towards a more comprehensive picture of azimuthal asymmetries within a TMD approach, touching upon at the same time universality, factorization and evolution issues. A corresponding study, along the same lines, for another important leading-twist, T-odd, TMD fragmentation function, the polarizing FF, was carried out in Ref.~\cite{DAlesio:2024ope}, by analyzing the transverse polarization of $\Lambda$ hyperons within a jet produced in unpolarized $pp$ collisions. 

As we will discuss below, the kinematics of this process covers regions in $z$, the pion longitudinal momentum fraction, and $x$, the light-cone momentum fraction of the quark in the polarized proton, complementary to those explored in SIDIS and $e^+e^-$ processes. In this respect, it can help to better constrain both the Collins FF and the transversity distribution in these regions.

In the spirit of a unified approach one might consider as well single spin asymmetries (SSAs or $A_N$) for inclusive pion production in $p^\uparrow p$ collisions, where only one large scale, the transverse momentum of the final-state hadron, is present. 
In such a case it is argued that the so-called collinear twist-three (CT3) formalism~\cite{Qiu:1991pp,Qiu:1998ia} is appropriate for their description. A global analysis adopting simultaneously the TMD (for two-scale processes) and the CT3 (for single-scale ones) approaches, and including also information from lattice QCD simulations, has been indeed successfully carried out in Ref.~\cite{Gamberg:2022kdb}.
Nevertheless, TMD and CT3 formalisms are intimately connected~\cite{Ji:2006ub}, and TMD and twist-three functions can be related (as also done in \cite{Gamberg:2022kdb}). 
In this regard, a recent study~\cite{Boglione:2024dal} has shown, for the first time, that a simultaneous analysis of the available experimental data for azimuthal asymmetries in SIDIS and $e^+e^-$ scattering processes, and SSAs in proton-proton collisions, assuming factorized expressions in terms of TMDs for all those processes, is possible. 
It is therefore worth exploring the impact of this simultaneous analysis in describing the Collins asymmetry for $p^\uparrow p\to {\rm jet}\; \pi\, X$ \footnote{For a recent attempt to include in the global fit of Ref.~\cite{Gamberg:2022kdb} data for pion-in-jet production in $p^\uparrow p$ collisions see Ref.~\cite{JAM2025}.}. 

The paper is organized as follows: in Section~\ref{sec:formalism} we recall the basic formulae of the theoretical formalism, then in Section~\ref{sec:results} we present our predictions and compare them against STAR data. In Section~\ref{sec:rew-pred} we explore the possibility to have an overall picture in terms of TMDs by adopting the results obtained in the simultaneous analysis performed in Ref.~\cite{Boglione:2024dal}, then in Section~\ref{sec:conclusions} we draw our conclusions.

\section{\label{sec:formalism} Formalism}

We recall here the basic elements of the formalism, providing the main expressions. All details can be found in Refs.~\cite{DAlesio:2010sag, DAlesio:2017bvu}.

We compute the azimuthal asymmetry for an unpolarized hadron $h$ produced within a jet in single-polarized hadron-hadron ($AB$) collisions, 
\begin{equation}
A^\uparrow (p_A)\,B(p_B)\rightarrow {\rm jet}(p_{\rm{j}})\,  h(p_h)\,X\, ,
\end{equation}
where $p_A,p_B, p_{\rm j}, p_h$ are the momenta of the incoming hadrons (two protons in our case), the jet, and the produced unpolarized hadron (a pion in the following) respectively.

We will consider transverse momentum effects only in the fragmentation mechanism within a leading order (LO) factorization framework, employing a collinear configuration for the initial state. While still allowing for the observed azimuthal modulation, this simplifies the treatment and gets closer to a TMD factorization picture where two measurable and ordered scales appear: the large transverse momentum of the jet and the small transverse momentum of the pion with respect to the jet. A similar approach has been successfully adopted in Ref.~\cite{DAlesio:2024ope}. 

For better clarity, we summarize below the kinematics adopted. We will work in the $AB$ center-of-mass (c.m.) frame, with $AB$ along the $z$ axis and the jet laying in the $xz$ plane.
The four-momenta (for massless hadrons and partons\footnote{The elementary process under consideration is $a(p_a)b(p_b)\to c(p_c) d(p_d)$.}) are given by
\begin{equation}
\begin{aligned}
&p_A^\mu = \frac{\sqrt s}{2}(1,0,0,1), \quad p_B^\mu  = \frac{\sqrt s}{2}(1,0,0,-1),\\
&p_a^\mu  =x_a \frac{\sqrt s}{2}(1,0,0,1), \quad p_b^\mu  =x_b \frac{\sqrt s}{2}(1,0,0,-1),  \\
&p_{\rm j}^\mu = E_{\rm j} (1, \sin\theta_{\rm j}, 0,\cos\theta_{\rm j}) = \pjT (\cosh\eta_{\rm j},1,0,\sinh\eta_{\rm j})\,,
\end{aligned}
\end{equation}
where $s$ is the c.m.~energy squared, $\eta_{\rm j}=-\log[\tan (\theta_{\rm j}/2)]$, is the jet pseudorapidity and $\pjT\equiv |\bm{p}_{{\rm j}T}|$ its transverse momentum in the $AB$ c.m.~frame.

The Collins contribution to the asymmetry for the $p^\uparrow p\to {\rm jet}\,\pi\, X$ process can be extracted by computing  
the following azimuthal moment:
\begin{equation}
\begin{aligned}
& A_N^{\sin(\phi_S-\phi^H_\pi)}(\bm{p}_{\rm j},z,p_{\perp \pi}) \\
= & 2\,\frac{\int\,d\phi_S \,d\phi_{\pi}^H\,\sin(\phi_S-\phi_\pi^H)\,[\,d\sigma(\phi_S,\phi_\pi^H)-
d\sigma(\phi_S+\pi,\phi_\pi^H)\,]} {\int\,d\phi_S\, d\phi_{\pi}^H\,[\,d\sigma(\phi_S,\phi_\pi^H)+d\sigma(\phi_S+\pi,\phi_\pi^H)\,]}\,,
\end{aligned}
\label{eq:an-col}
\end{equation}
where $d\sigma(\phi_S,\phi_\pi^H)$ stands for the  differential cross section
\begin{equation}
\frac{E_{\rm j}\,d\sigma^{p(S,\phi_S) p\to {\rm jet}\,\pi(\phi_\pi^H)\,X}}
{d^3\bm{p}_{\rm j}\,dz\,d^2\bm{p}_{\perp\pi}\,\,d\phi_S}\,.
\label{eq:dsig}
\end{equation}
Here $z$ is the pion longitudinal momentum fraction\footnote{Focusing on light mesons, we will identify the light-cone momentum fraction with the longitudinal momentum fraction.}, $\bm{p}_{\perp\pi}\, \equiv \bm{j}_T$ is the transverse momentum of the pion with respect to the parent parton, $\phi_\pi^H$ is the azimuthal angle of the pion momentum, measured in the jet helicity frame, and $S$ is the (transverse) polarization vector of the initial proton beam, forming an angle $\phi_S$ with the jet production plane (the $xz$ plane) in the $pp$ c.m.~reference frame.
The azimuthal factor $\sin(\phi_S-\phi^H_\pi)$ in the numerator of Eq.~(\ref{eq:an-col}) singles out the Collins contribution.

The numerator in Eq.~(\ref{eq:an-col}) can be schematically given as
\begin{equation}
N[A_N^{\sin(\phi_S-\phi_\pi^H)}] \sim\!\sum_{a,b,c,d} h_1^a(x_a) \otimes
 f_1^b(x_b)\otimes \Delta\hat\sigma^{ab\to cd} \otimes H_1^{\perp\, c}(z,\bm{p}_{\perp \pi}^2)\,,
\label{eq:Coll2}
\end{equation}
that is a convolution of the {\em collinear} quark transversity distribution, $h_1^a$, with the Collins TMD FF, $H_1^{\perp c}$, and the partonic spin transfer, $\Delta\hat\sigma$, while the denominator is simply twice the unpolarized cross section, where all partons, quarks and gluons, are included. The scale dependence of the nonperturbative functions 
in Eq.~(\ref{eq:Coll2}) has been understood.  
We will comment on this below. 
Note that, within a LO 
approach, the final 
parton $c$  in the 
elementary process $ab \to cd$ (a quark in the present case) can be identified with the observed fragmenting jet: $c\equiv $ jet. 

For the $x$-dependent part of the transversity function we adopt the following parametrization at the initial scale~\cite{Anselmino:2007fs, Anselmino:2013vqa, Anselmino:2015sxa, DAlesio:2020vtw}: 
\begin{equation}\label{eq:h1(x)-SB}
 h_1^q(x\,, Q_0^2) = {\mathcal N}^T_q(x) \frac12 \left[f_{q/p}(x\,, Q_0^2) + g_{1L}^q(x\,, Q_0^2)\right]\,,
\end{equation}
where
\begin{equation}
 {\cal N}^{T}_q(x)=N^{T}_q x^{\alpha}(1-x)^\beta\,
\frac{(\alpha+\beta)^{\alpha+\beta}}{\alpha^\alpha \beta^\beta},
\quad (q = u_v,\,d_v)\,,
\end{equation}
with the $\alpha$ and $\beta$ parameters as extracted in Ref.~\cite{Boglione:2024dal}. Notice that we consider only the valence contributions. We set $Q_0^2 = 0.81\,\text{GeV}^2$ as the input scale, with $\alpha_S(M_Z) \simeq 0.118$.

The Collins functions are parametrized as in Refs.~\cite{Anselmino:2007fs, Anselmino:2013vqa, Anselmino:2015sxa, DAlesio:2020vtw}:
\begin{equation}
\label{eq:Collins}
H_1^{\perp q}(z, p_\perp^2)  = {\cal N}^C_q (z) \frac{z m_h}{ M_C}\,\sqrt{2e}\,e^{-p_\perp^2/M_C^2}\,D_{h/q}(z, p_\perp^2)\,,
\end{equation}
where $q = \rm{fav}, \rm{unf}$ (favored/unfavored fragmentation functions), $m_h$ is the produced hadron mass and $p_\perp = |\bm{p}_\perp|$. $D_{h/q}(z, p_\perp^2)$ is the unpolarized TMD FF, while the ${\cal N}^C_q(z)$ factors are given by
\begin{equation}
\label{eq:Collins-NC}
 {\cal N}^C_{\rm{fav}}(z) = N^C_{\rm{fav}}\, z^\gamma, 
 \quad{\cal N}^C_{\rm{unf}}(z) = N^C_{\rm{unf}}\,.
\end{equation}
All these parameters have been extracted in Ref.~\cite{Boglione:2024dal} by fitting data on SIDIS and $e^+e^-$ azimuthal asymmetries. 

The unpolarized TMD FFs are  parametrized using a factorized Gaussian ansatz:
\begin{equation}
\begin{aligned}
 & D_{h/q}(z, p_\perp^2) = D_{h/q}(z)\, \frac{e^{-p^2_\perp / \langle p^2_\perp \rangle}}{\pi \langle p^2_\perp\rangle}\,,
\end{aligned}
\end{equation}
with $\langle p_\perp^2\rangle = 0.12$ GeV$^2$ as extracted from a fit to HERMES multiplicities~\cite{Anselmino:2013lza}. As for the collinear parts, we adopt the MSHT20nlo proton PDFs~\cite{Bailey:2020ooq} and the DEHSS fragmentation functions for pions~\cite{deFlorian:2014xna,deFlorian:2017lwf}.  
Apart from the transversity function in Eq.~\eqref{eq:h1(x)-SB}, that is evolved to higher $Q^2$ values through a modified version of {\tt HOPPET}~\cite{Salam:2008qg,Prokudin:hoppet}, in all cases, including the collinear part of the Collins function (see Eq.~(\ref{eq:Collins})), the scale dependence is computed via DGLAP evolution equations. 

Before presenting our results, a few comments are in order.
Let us start with the choice of the factorization scale, both in the collinear PDFs and in the TMD FFs. On the one hand, 
the relevant scale for the TMD FFs for this kind of processes is $\mu_j=\pjT R$, where $R$ is the jet-cone radius, as discussed in Ref.~\cite{Kang:2017glf}. Then, by evolving up to $\mu=p_{{\rm j}T}$, one resums single logarithms in the jet size parameter to all orders in the strong coupling constant $\alpha_s$. On the other hand, working at LO accuracy, which is independent from $R$ (and more generally from the jet dynamics), and following 
Ref.~\cite{Kang:2017btw}, we can safely use $\mu=\pjT$ for the unpolarized and the Collins TMD FFs and similarly for the collinear distributions.

Another relevant issue is that, in order to determine the proper partonic frame, one should consider the backward going jet together with the jet around which we measure the final hadron, as pointed out in Ref.~\cite{Boer:2007nh}. Indeed, the $z$ and $p_\perp$ dependences of the FF for a jet with a large transverse momentum are not invariant under Lorentz boosts along the $pp$ direction. This could cause a mismatch with the corresponding variables adopted in SIDIS and $e^+e^-$ expressions. Since these differences scale as $E_{\rm jet}/\sqrt s$~\cite{Boer:2010yp}, we expect that, for the energies and the kinematics considered here, these corrections are not larger than a few percent. 

\section{\label{sec:results} Results}

We present here our theoretical estimates of the Collins azimuthal asymmetry for $p^\uparrow p\to {\rm jet}\, \pi\, X$, and  compare them against STAR data~\cite{STAR:2022hqg,Zhang:2024zuq,STAR:2025xyp}, adopting the corresponding kinematics and experimental setups. In particular, we consider data in the forward region\footnote{All theoretical estimates are obtained by imposing proper cuts on the cone radius as adopted in the experimental analyses.}, that is $\eta_{\rm j} \ge 0$, and all our estimates are integrated in the range $0\le \eta_{\rm j}\le 0.9$.

Our results are based on the latest extractions performed in Ref.~\cite{Boglione:2024dal}, where, with the final aim to include also a new set of data within a simultaneous reweighting procedure, we reanalyzed SIDIS and $e^+e^-$ azimuthal asymmetries.
More precisely, we adopt the updated extraction of $h_1^q$ and $H_1^{\perp q}$ from Ref.~\cite{Boglione:2024dal}, that was used as a prior for the Bayesian reweighting. The predictions from the reweighted TMDs, including the information from SSA data in inclusive processes, will be presented separately in Section~\ref{sec:rew-pred}. 

In what follows, we always adopt the median of the Monte Carlo sets (generated to  estimate the uncertainty on the extraction) as central value, and the uncertainties are  computed at a $2\sigma$ confidence level (CL). 

It is worth stressing  that, as already discussed in  Ref.~\cite{DAlesio:2017bvu}, these are genuine predictions based on extractions from processes for which TMD factorization has been rigorously proven. In this respect the present study can be considered a further indication of the possible and somehow expected extension of TMD factorization also for pion-in-jet production in $pp$ collisions.

\begin{figure}[!t]
\centering
\includegraphics[width=6.75cm, keepaspectratio]{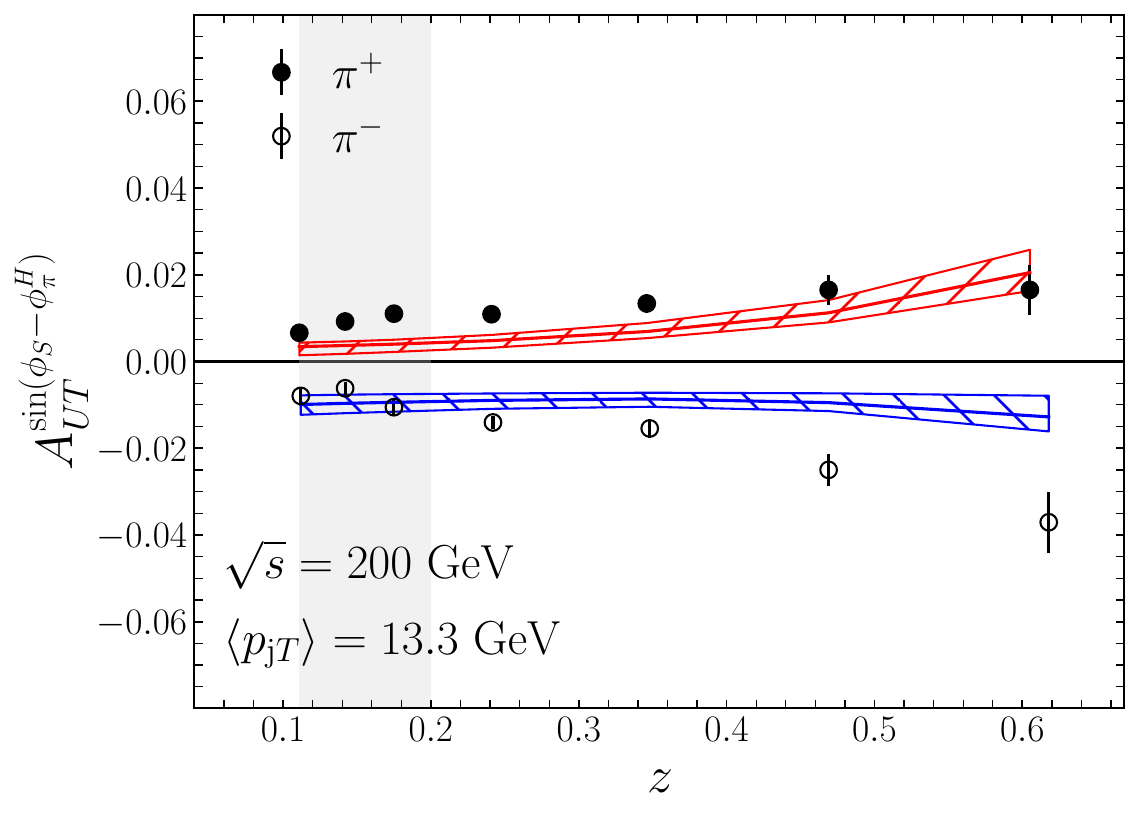}\\
\includegraphics[width=6.75cm, keepaspectratio]{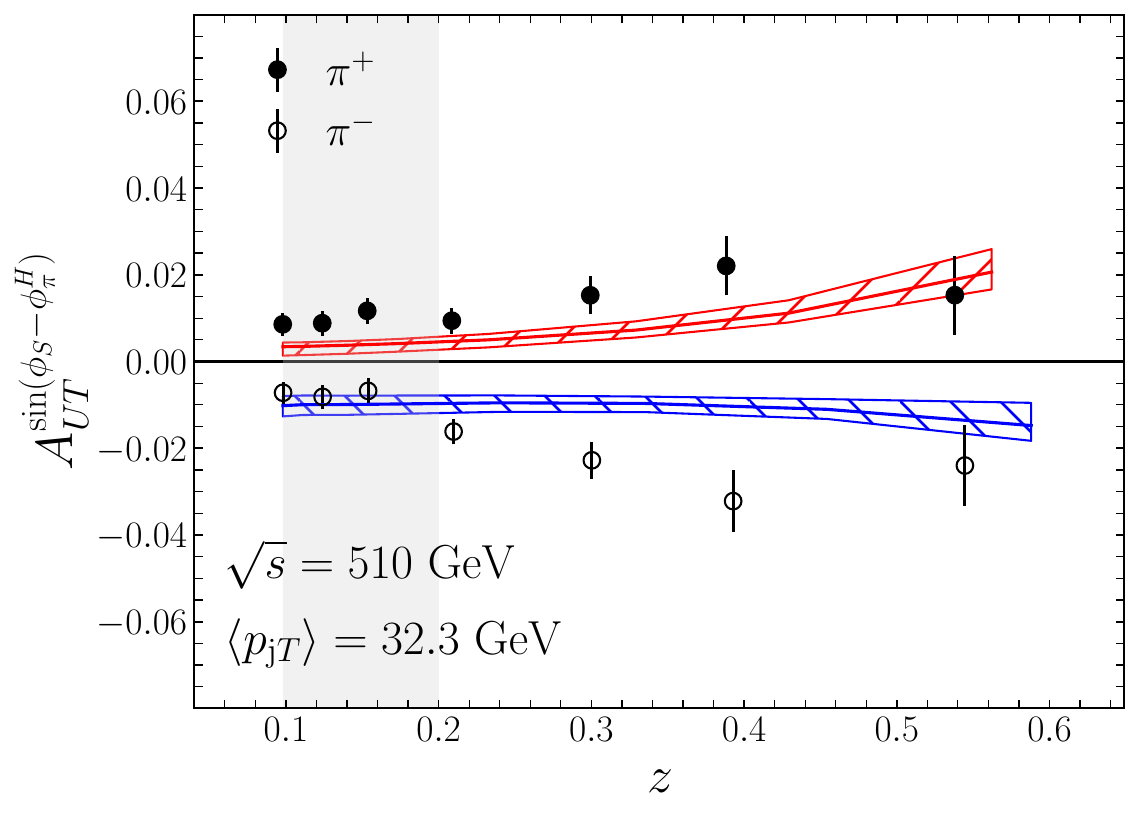}
\caption{
Estimates of the Collins azimuthal asymmetry for $p^\uparrow p\to {\rm jet}\, \pi\, X$, based on the extractions of Ref.~\cite{Boglione:2024dal}, as a function of $z$, integrated over $\eta_{\rm j}$ in the forward region, at $\sqrt s=200$~GeV and $\langle \pjT\rangle = 13.3$~GeV (upper panel) and  $\sqrt s= 510$~GeV and $\langle \pjT\rangle = 32.3$~GeV (lower panel). Uncertainty bands at 2$\sigma$ CL are also shown. STAR data are from Refs.~\cite{STAR:2022hqg,
STAR:2025xyp}.
}
\label{fig:asymz} 
\end{figure}

\begin{figure}[!h]
\centering
\includegraphics[width=6.75cm, keepaspectratio]{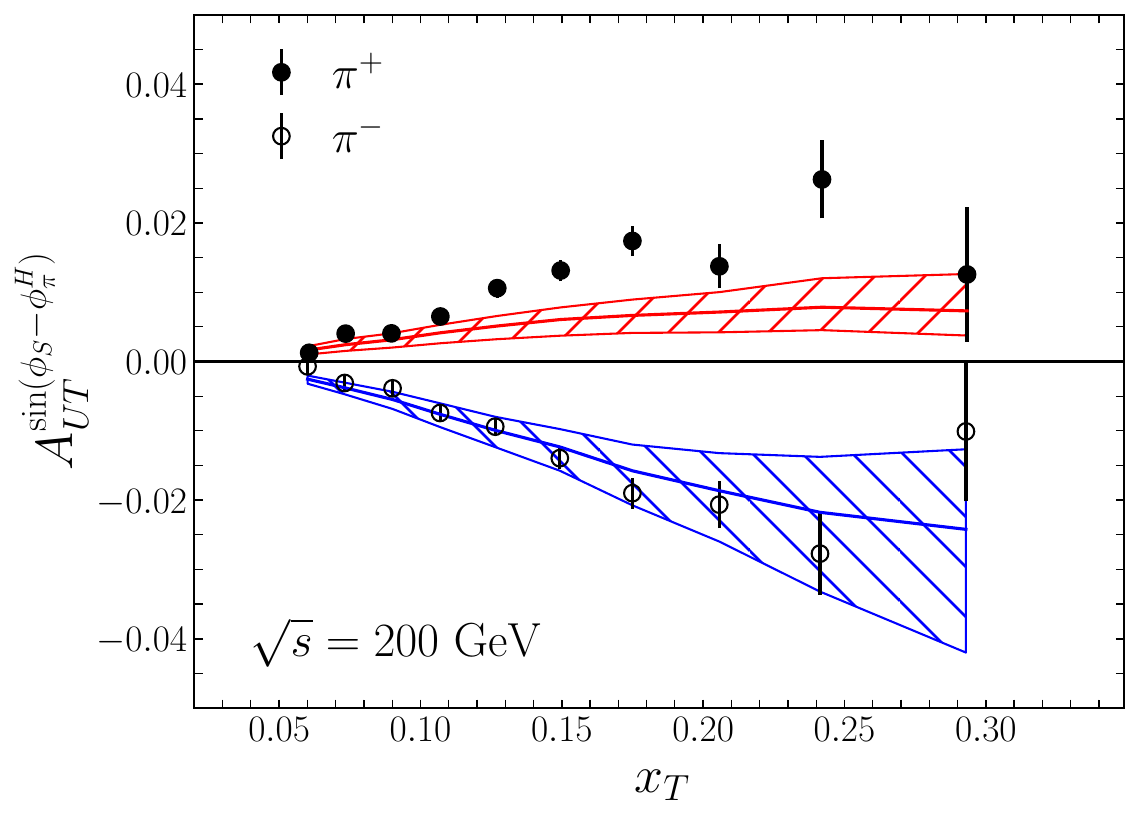}\\
\includegraphics[width=6.75cm, keepaspectratio]{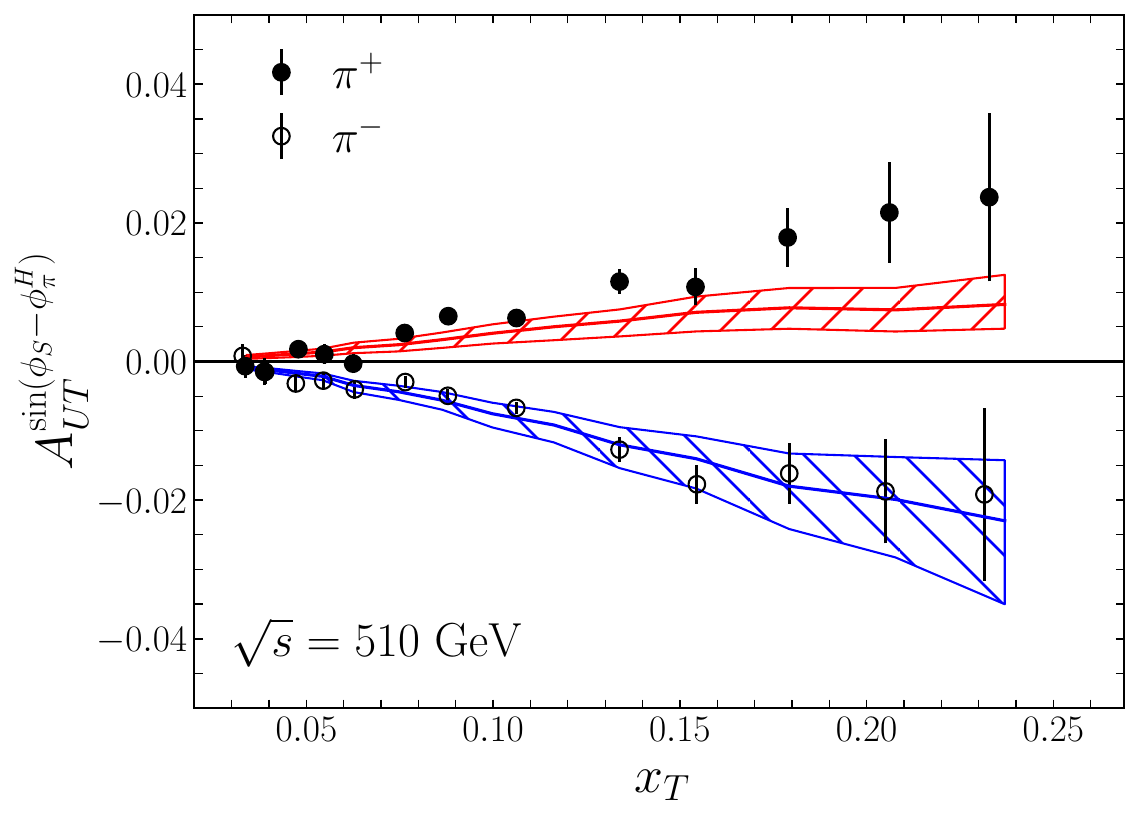}
\caption{Estimates of the Collins azimuthal asymmetry for $p^\uparrow p\to {\rm jet}\, \pi\, X$, based on the extractions of Ref.~\cite{Boglione:2024dal}, as a function of $x_T$ at $\sqrt s=200$~GeV (upper panel) and $\sqrt s= 510$~GeV (lower panel), integrated over $z$ and over $\eta_{\rm j}$ in the forward region. Uncertainty bands at 2$\sigma$ CL are also shown. STAR data are from Refs.~\cite{STAR:2022hqg, STAR:2025xyp}.
}
\label{fig:asymxt} 
\end{figure}
\begin{figure}[!h]
\centering
\includegraphics[width=\linewidth, keepaspectratio]{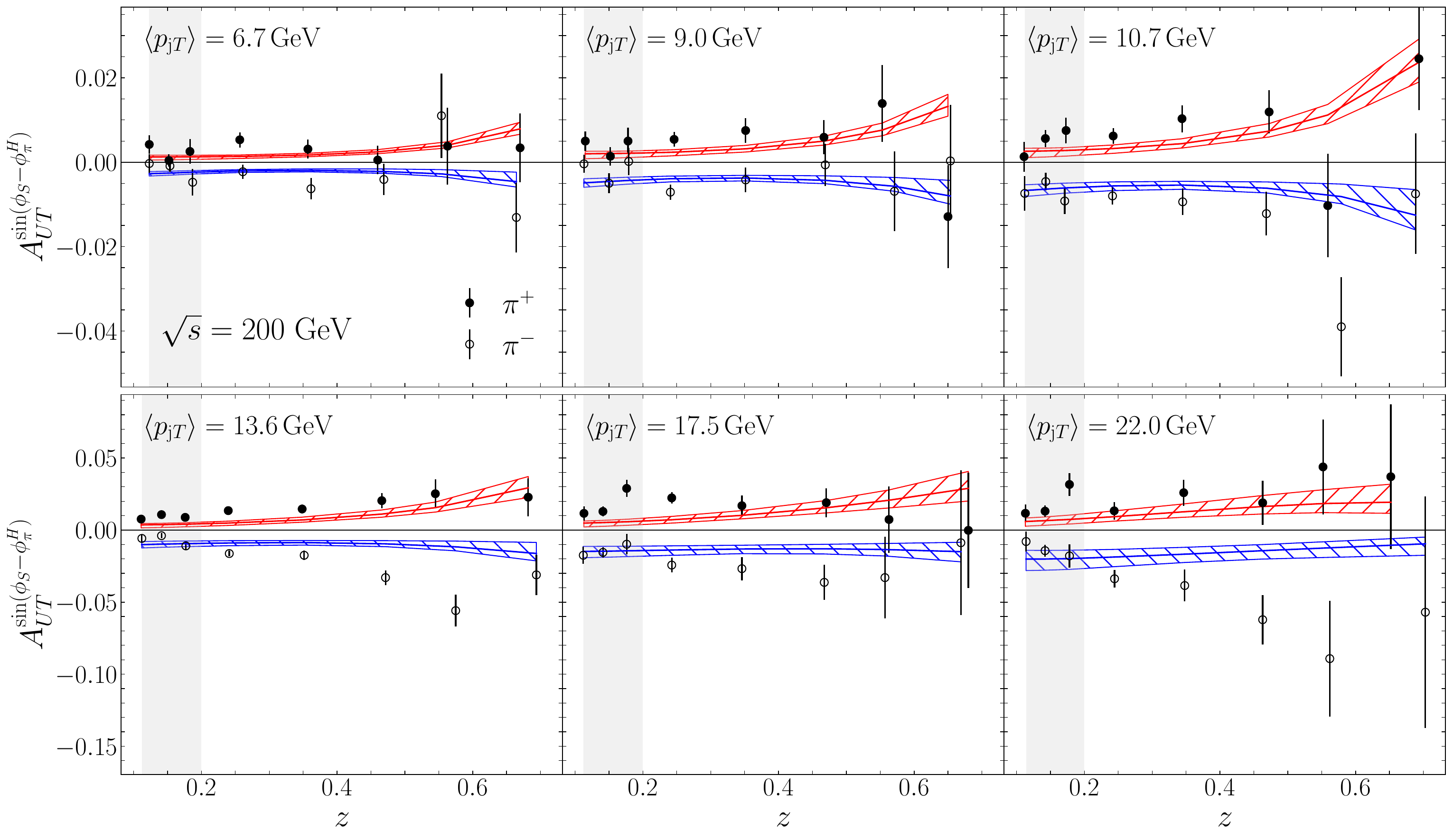}\\
\includegraphics[width=\linewidth, keepaspectratio]{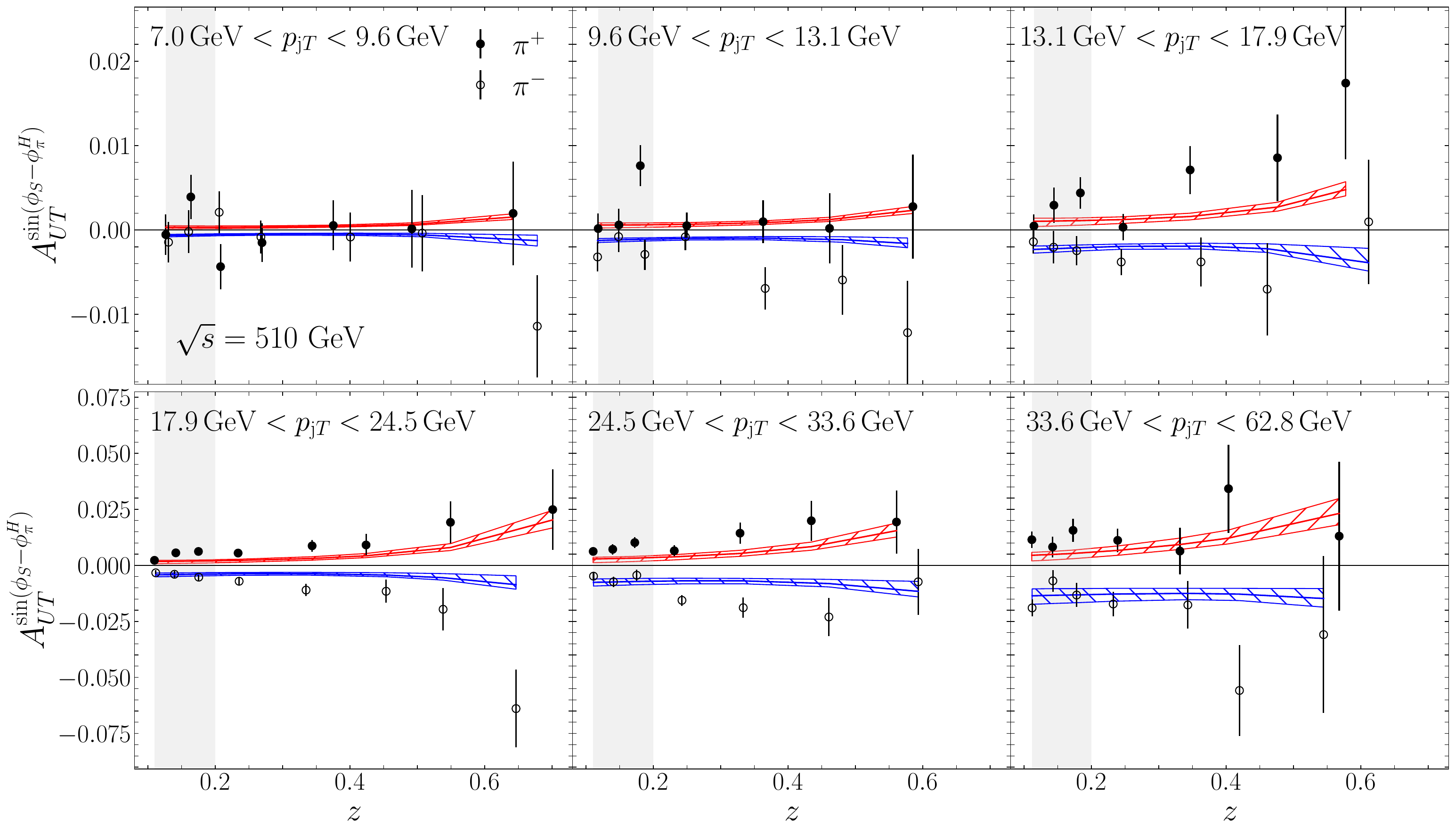}
\caption{Estimates of the Collins azimuthal asymmetry for $p^\uparrow p\to {\rm jet}\, \pi\, X$, based on the extractions of Ref.~\cite{Boglione:2024dal}, as a function of $z$ at $\sqrt s= 200$~GeV (upper panels) and  $\sqrt s= 510$~GeV (lower panels) integrated over $\eta_{\rm j}$ in the forward region, for different $\pjT$ bins. Uncertainty bands at 2$\sigma$ CL are also shown. STAR data are from Refs.~\cite{STAR:2022hqg, Zhang:2024zuq}.
}
\label{fig:asymzbins} 
\end{figure}

In Fig.~\ref{fig:asymz} we show our estimates for the Collins azimuthal asymmetry for charged pions at $\sqrt s= 200$ GeV and $\langle \pjT\rangle = 13.3$~GeV (upper panel) and at $\sqrt s= 510$~GeV  and $\langle \pjT\rangle = 32.3$~GeV (lower panel) as a function of $z$, while in Fig.~\ref{fig:asymxt} we present the corresponding ones as a function of $x_T = 2\,\pjT/\sqrt s$, integrated over $z$ in the range $[0.1\!:\!0.8]$. Data from STAR at 200 GeV~\cite{STAR:2022hqg} and 510 GeV~\cite{STAR:2025xyp} are also displayed. Moreover, to provide a more detailed picture, in Fig.~\ref{fig:asymzbins} we give our predictions for $A_N^{\sin(\phi_S-\phi^H_\pi)}$ as a function of $z$ at $\sqrt s= 200$~GeV \cite{STAR:2022hqg} (upper panels) and $510$~GeV \cite{Zhang:2024zuq} (lower panels) for different $\pjT$ bins. 

The overall agreement with STAR data is definitely good and no significant energy dependence appears. This supports once again the conjectured validity of TMD factorization for these nonstandard TMD processes showing, at the same time, that any evolution effect is milder in such observables (see below). 

Only some discrepancy, mostly for $\pi^-$ data at moderate and large $z$, is observed in Fig.~\ref{fig:asymz} and Fig.~\ref{fig:asymzbins}. This can be traced back to the relatively simple parametrization adopted for the Collins functions that, for the unfavored one, assumes the same $z$ dependence as the unpolarized TMD FF (see Eq.~\eqref{eq:Collins-NC}). More flexible parametrizations are highly desirable and deserve extra analyses beyond the scope of this work. We have also to remark that the extraction of the Collins function from $e^+e^-$ data is limited to the region $z> 0.2$. This means that we are entering an unexplored region. For these reasons in Fig.~\ref{fig:asymz} we have suitably indicated this by shadowing the region below $z=0.2$.

A word of caution is nevertheless mandatory. Indeed, one cannot exclude that, within this simplified LO approach, a sort of cancellation between higher-order corrections and potential factorization-breaking contributions might occur. On the other hand, the general agreement between data and LO predictions is quite encouraging. Further improved studies along this direction, as well as more experimental analyses, will certainly help in shedding light on these issues.

Concerning the $x_T$ dependence, we have to recall that at fixed $x_T$ we are probing ({\it i.e.}~integrating over) the transversity distributions at $x>x_T$, that is we are entering a region poorly explored by SIDIS data ($x\gtrsim 0.3$). In this respect, the discrepancy in size for $\pi^+$ production, at moderate $x_T$ values, can be traced back to the poorly known behavior of the up-quark transversity distribution at large $x$. Once again, a combined fit, including this data set, could help in constraining $h_1^q$ in this unexplored region. 

\begin{figure}[t]
\centering
\hspace*{-.5cm}
\includegraphics[width=4.5cm, keepaspectratio]{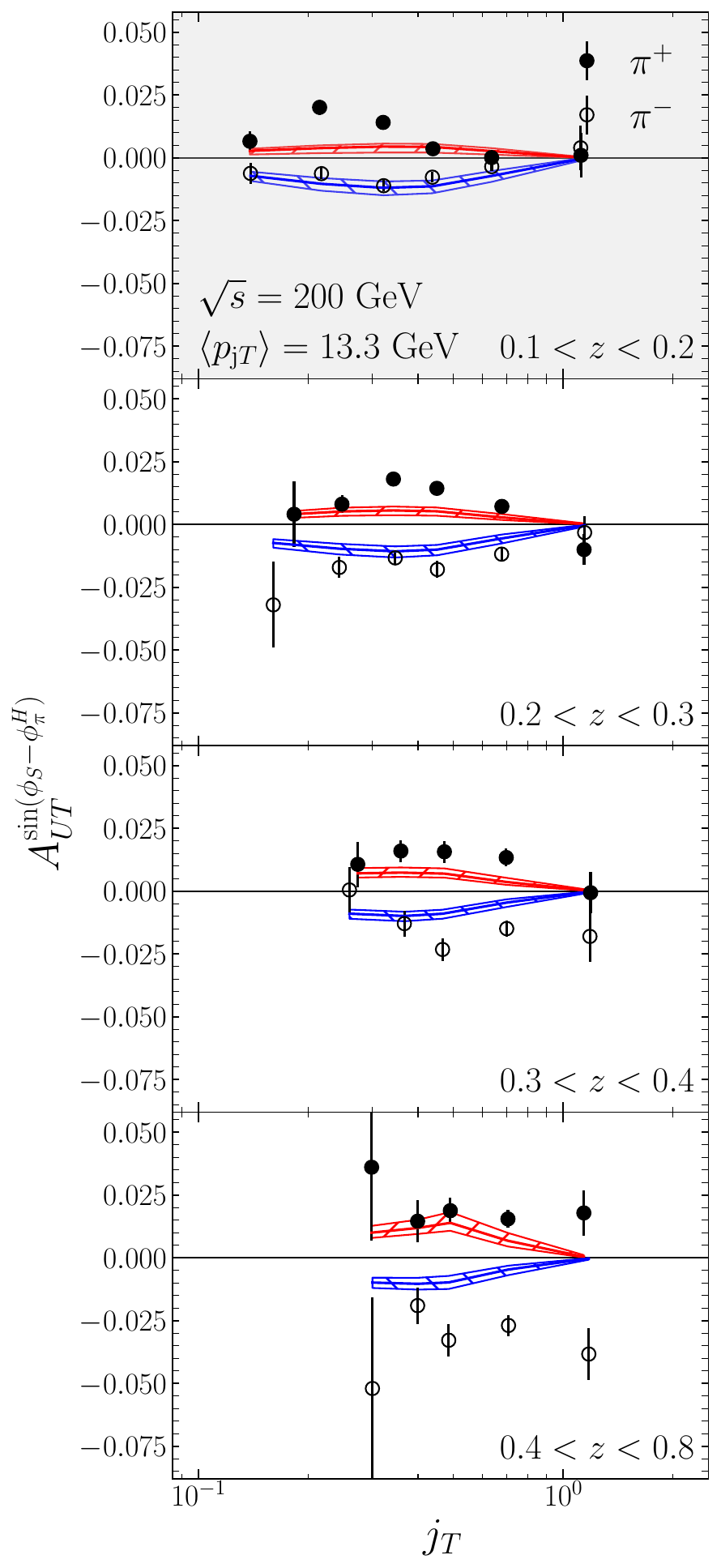}
\includegraphics[width=4.5cm, keepaspectratio]{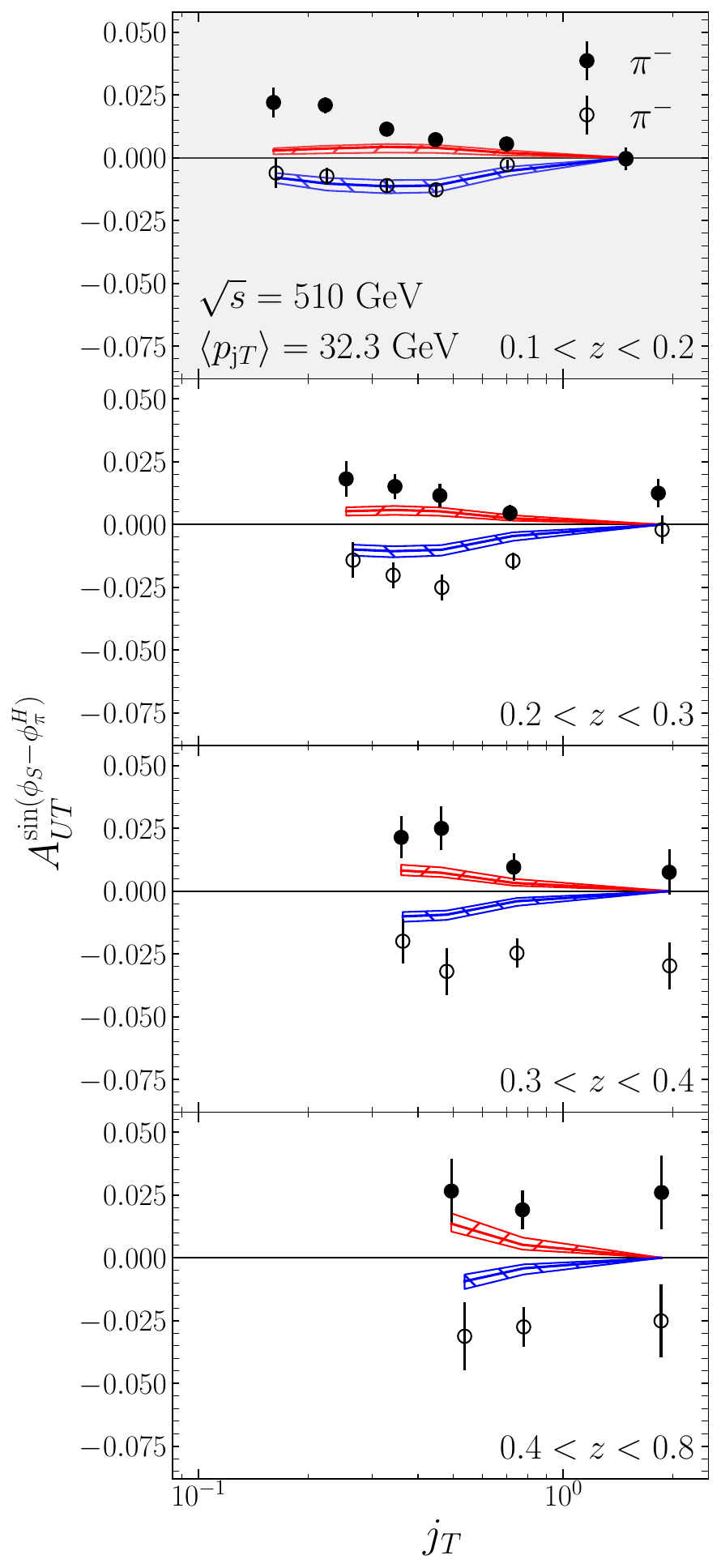}
\caption{Estimates of the Collins azimuthal asymmetry for $p^\uparrow p\to {\rm jet}\, \pi\, X$, based on the extractions of Ref.~\cite{Boglione:2024dal}, as a function of $j_T$, at $\sqrt s=200$~GeV (left panels) and $\sqrt s= 510$~GeV (right panels) integrated over $\eta_{\rm j}$ in the forward region, for different $z$ bins. In both cases we use the corresponding $\langle \pjT \rangle$ values. Uncertainty bands at 2$\sigma$ CL are also shown. STAR data are from Refs.~\cite{STAR:2022hqg,STAR:2025xyp}.
}
\label{fig:asymjt} 
\end{figure}

To further investigate these findings, in Fig.~\ref{fig:asymjt} we compare our predictions against STAR data at $\sqrt s = 200$~GeV \cite{STAR:2022hqg}  (left panels) and $\sqrt s= 510$~GeV \cite{STAR:2025xyp} (right panels) as a function of $j_T$ for different $z$ bins. These distributions are indeed very challenging and extremely useful to learn about the intrinsic transverse momentum in the fragmentation mechanism and its correlation with the hadron longitudinal momentum fraction. Moreover, they represent a tool to access the  relative $j_T$ dependence  between the Collins and the unpolarized TMD FFs.

On the other hand, we have to keep in mind that we are adopting a very simple parametrization of the $j_T$ distribution for the TMD FFs: the simplest one indeed, a Gaussian-like form, with no flavor and/or $z$ dependences in the Gaussian width. These two assumptions could be certainly reconsidered in future studies and the model can be further improved. Despite of this and except for some bins, the agreement is very good. Notice that the upper panels, corresponding to the unexplored $z$ region, are consistently shadowed. It is nevertheless worth stressing the excellent agreement for $\pi^-$ data at both energies. Concerning the larger $z$ values a more flexible parametrization could definitely help in reducing the residual discrepancies.

\begin{figure}[h!]
\centering
\includegraphics[width=\linewidth, keepaspectratio]{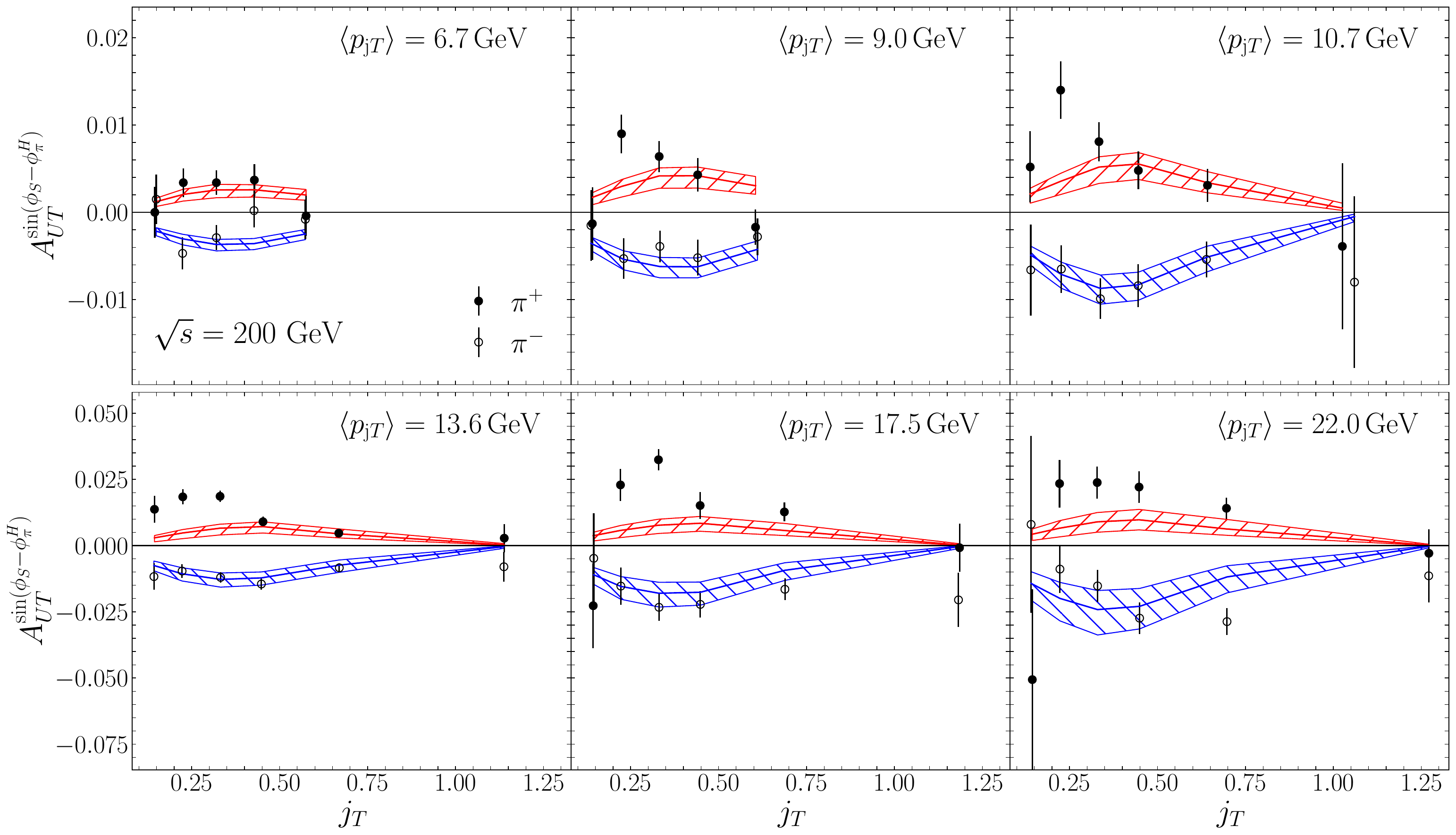}\\
\includegraphics[width=\linewidth, keepaspectratio]{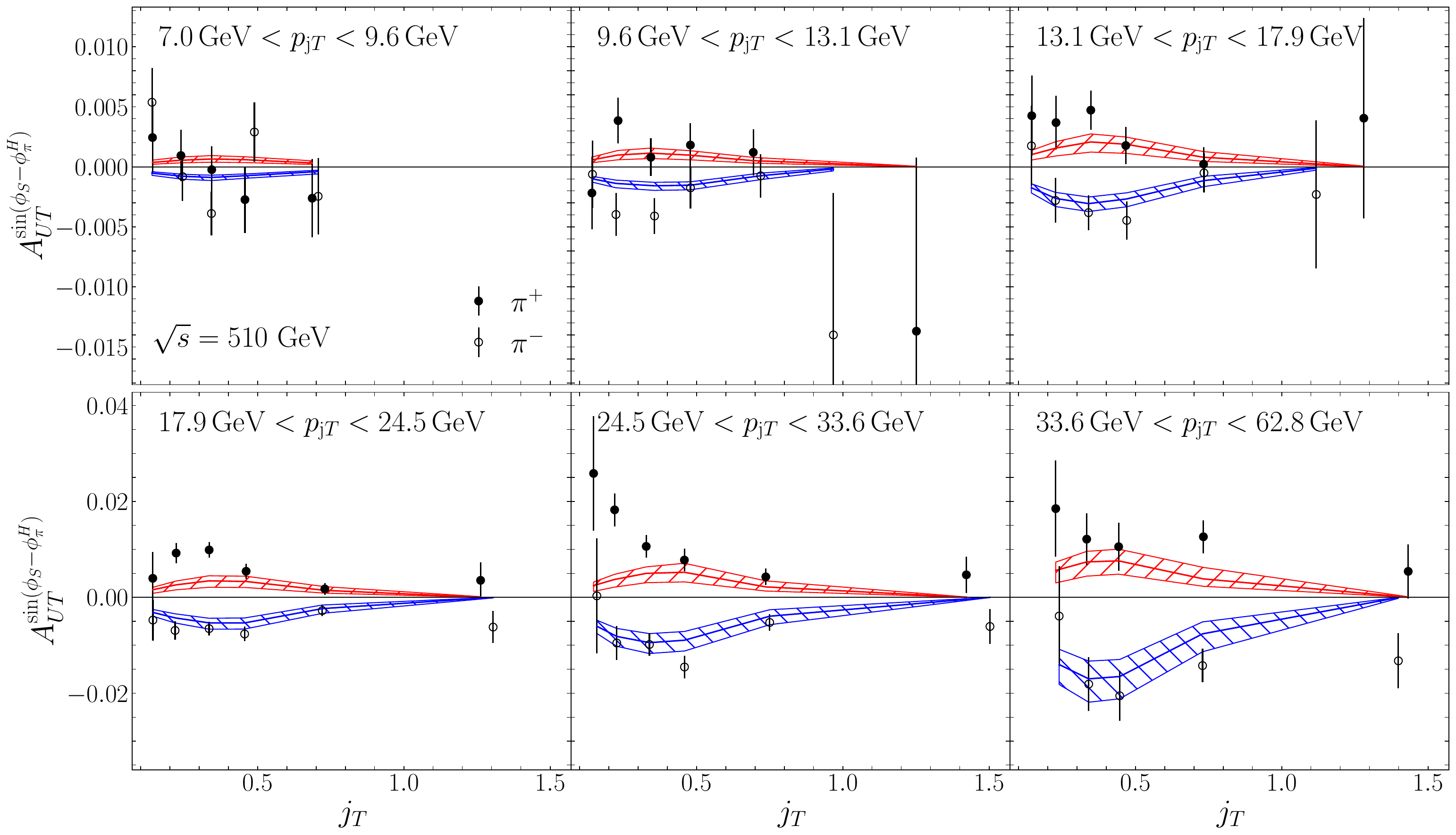}
\caption{Estimates of the Collins azimuthal asymmetry for $p^\uparrow p\to {\rm jet}\, \pi\, X$, based on the extractions of Ref.~\cite{Boglione:2024dal}, as a function of $j_T$ at $\sqrt s= 200$~GeV (upper panels) and $\sqrt s= 510$~GeV (lower panels) for different $\pjT$ bins, integrated over $z$ and over $\eta_{\rm j}$ in the forward region. Uncertainty bands at 2$\sigma$ CL are also shown. STAR data are from Refs.~\cite{STAR:2022hqg, Zhang:2024zuq}.
}
\label{fig:asymjtbins} 
\end{figure}

A similar situation holds for Fig.~\ref{fig:asymjtbins}, where we show our estimates against data at $\sqrt s = $ 200~GeV \cite{STAR:2022hqg} (upper panels) and $\sqrt s = $ 510~GeV \cite{Zhang:2024zuq} (lower panels) as a function of $j_T$ for different $\pjT$ bins, integrated over $z$.  This comparison is in favor of mild evolution effects with the hard scale even for the intrinsic transverse momentum dependence. Nonetheless, and once again, a more detailed and educated choice for the $j_T$ dependence in the TMD parameterizations and/or a more accurate treatment of the jet fragmentation mechanism~\cite{Kang:2020xyq,Kang:2021ffh, Kang:2023big, Bacchetta:2023njc} would certainly help in improving the description.

\subsection{\label{sec:rew-pred} Predictions from  reweighted extractions}

We finally present a comparison of a subset of predictions, still against STAR data, based on the reweighted $h_1^q$ and $H_1^{\perp q}$ extractions from Ref.~\cite{Boglione:2024dal}, where also SSA data for inclusive pion production in polarized $pp$ collisions were considered. The estimates are given for two effective approaches: the generalized parton model (GPM)~\cite{DAlesio:2004eso,Anselmino:2005sh,DAlesio:2007bjf} and its color-gauge invariant (CGI-GPM)~\cite{Gamberg:2010tj,DAlesio:2011kkm, DAlesio:2017rzj,DAlesio:2018rnv} extension. Though the latter approach affects only the Sivers contribution, the  transversity and Collins extractions are anyway differently weighted. In fact, in the computation of $A_N$ for polarized $pp$ collisions the Sivers contribution cannot be separated from the Collins effect and the latter turns to be slightly different in the GPM and in the CGI-GPM.

\begin{figure}[h!]
\centering
\includegraphics[width=4.35cm, keepaspectratio]{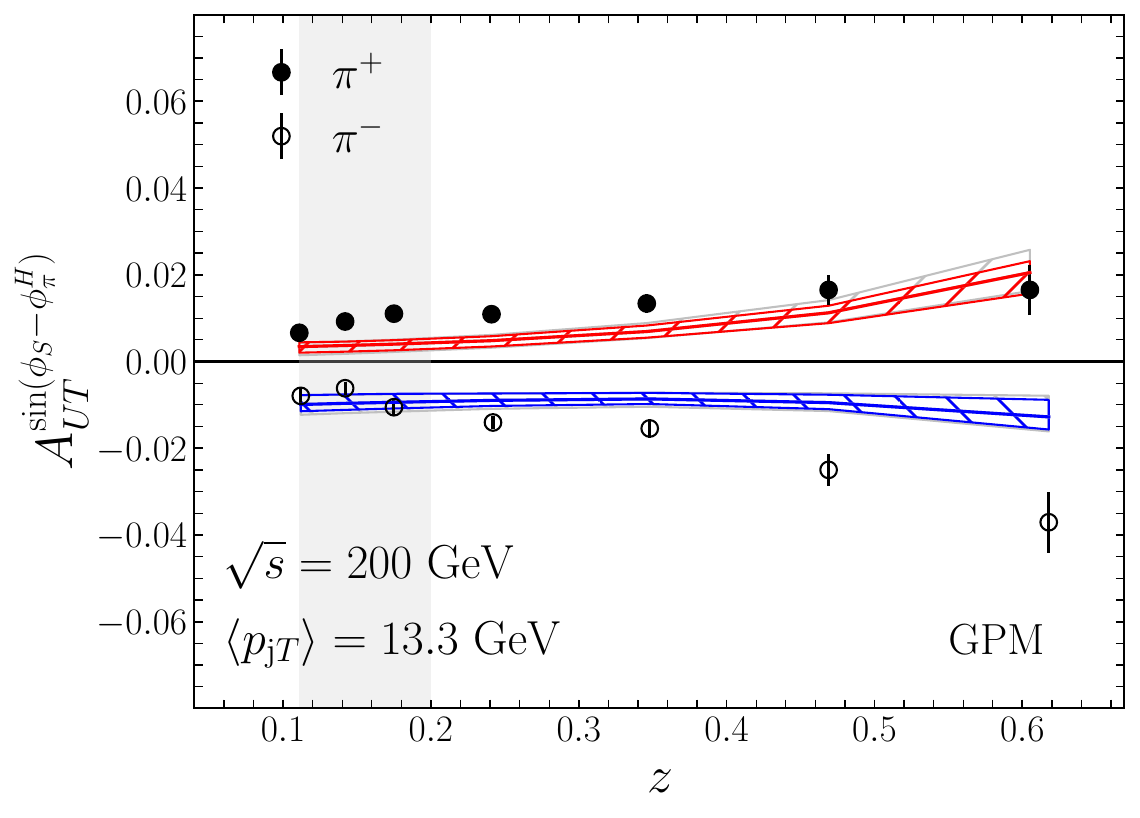}
\hspace*{-.1cm}
\includegraphics[width=4.35cm, keepaspectratio]{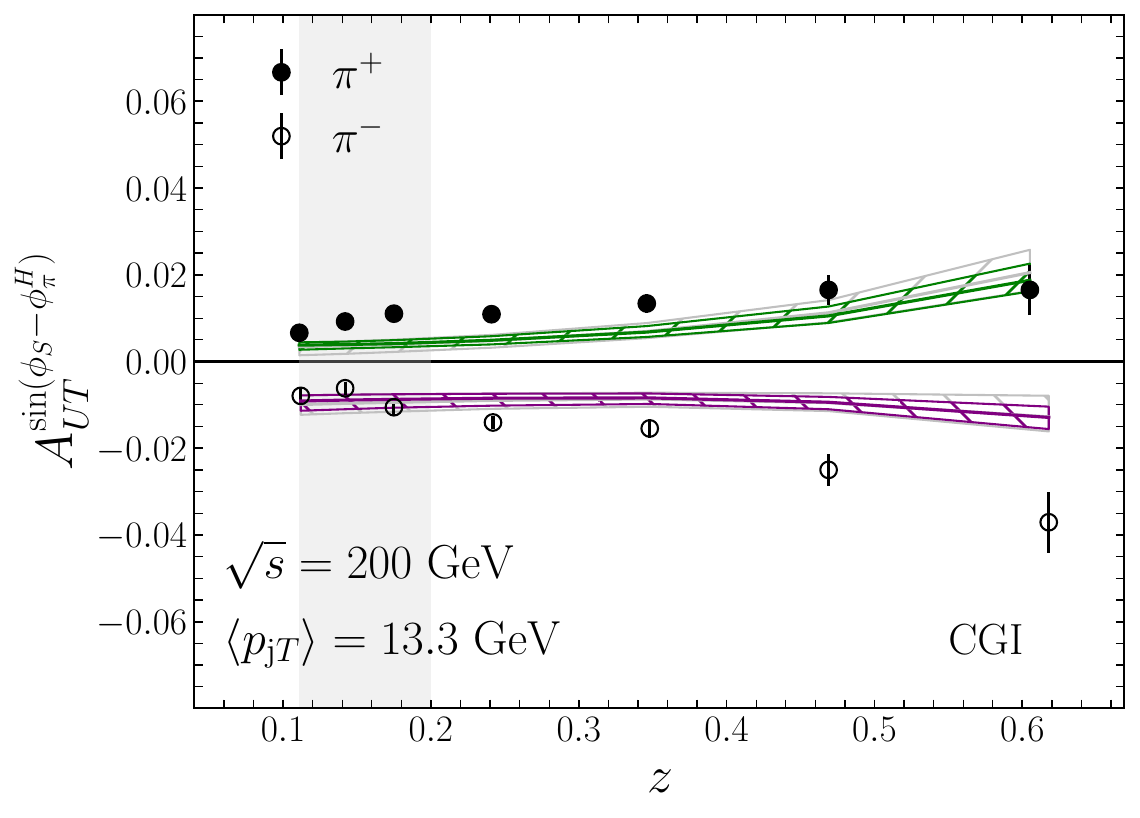}\\
\includegraphics[width=4.35cm, keepaspectratio]{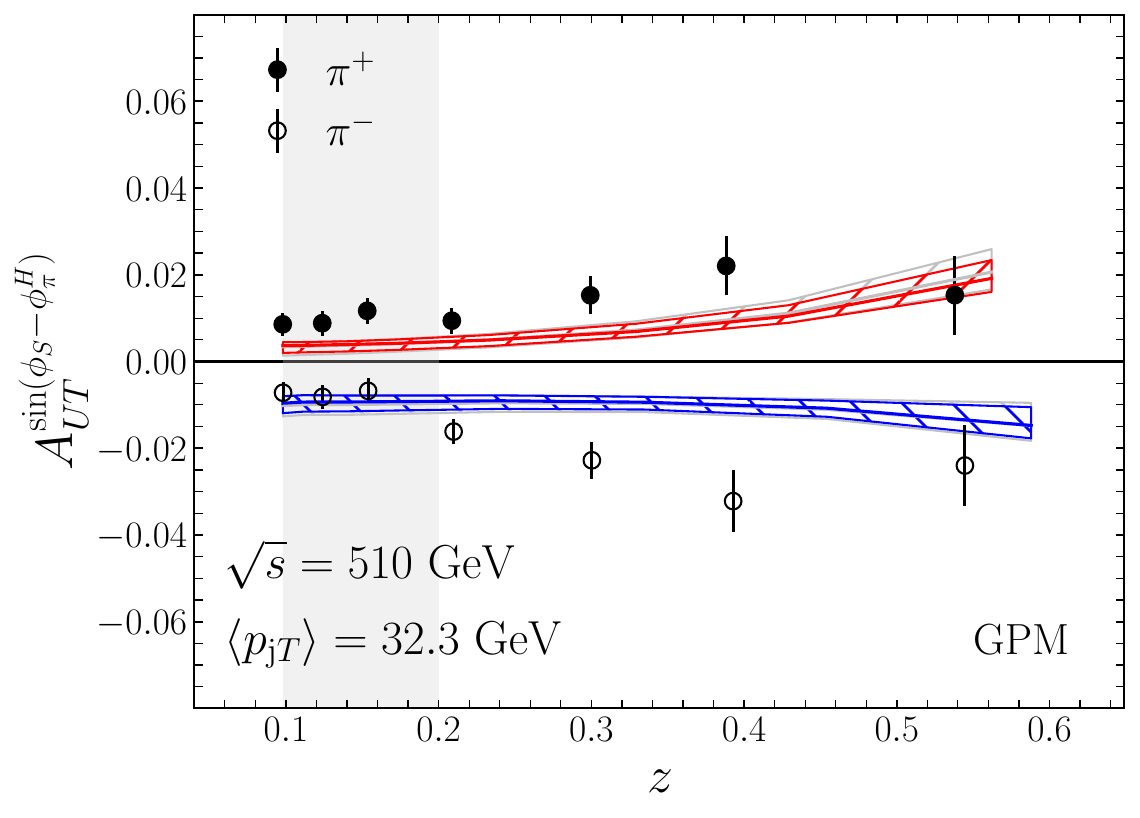}
\includegraphics[width=4.35cm, keepaspectratio]{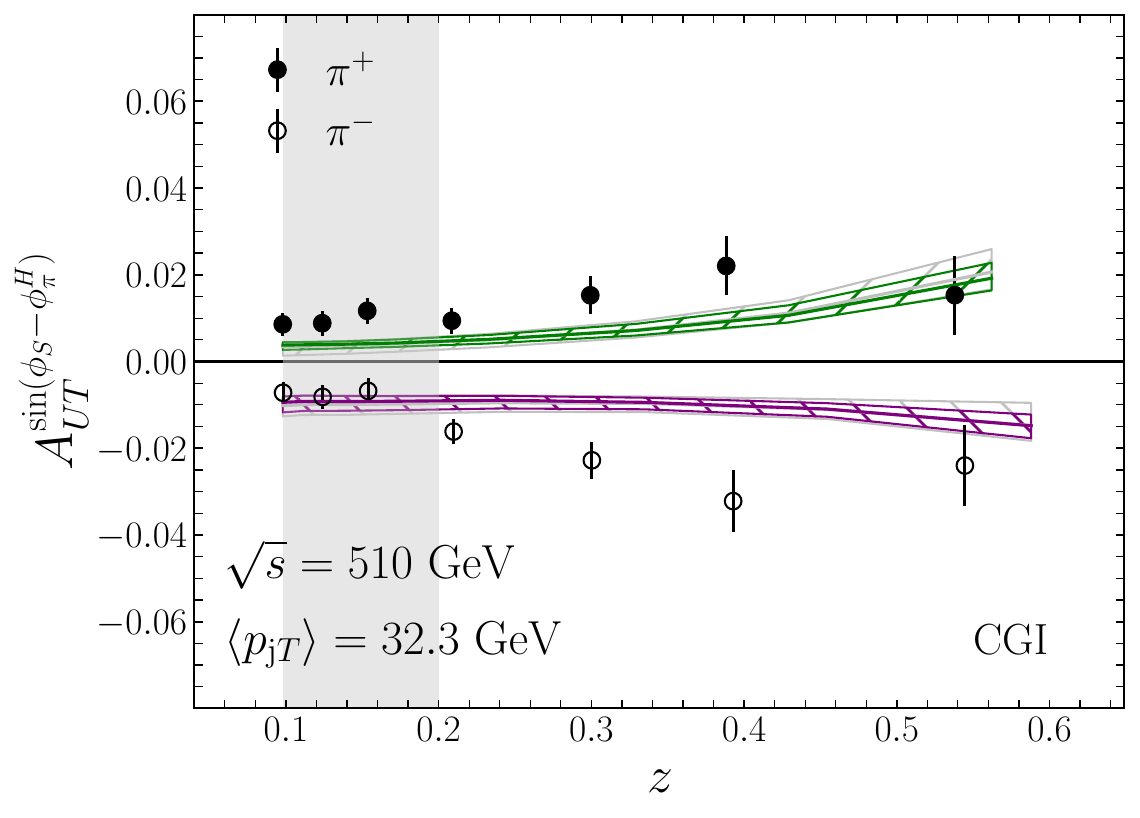}
\caption{Estimates of the Collins azimuthal asymmetry for $p^\uparrow p\to {\rm jet}\, \pi\, X$, based on the reweighted extractions of Ref.~\cite{Boglione:2024dal}, as a function of $z$ and integrated over $\eta_{\rm j}$ in the forward region, at $\sqrt s=200$~GeV and $\langle \pjT\rangle = 13.3$~GeV (upper panels) and at $\sqrt s= 510$~GeV and $\langle \pjT\rangle = 32.3$~GeV (lower panels) for the GPM (left panels) and CGI-GPM (right panels) formalisms. Uncertainty bands and data as in Fig.~\ref{fig:asymz}.
}
\label{fig:asymz-rew} 
\end{figure}

\begin{figure}[h!]
\centering
\includegraphics[width=4.35cm, keepaspectratio]{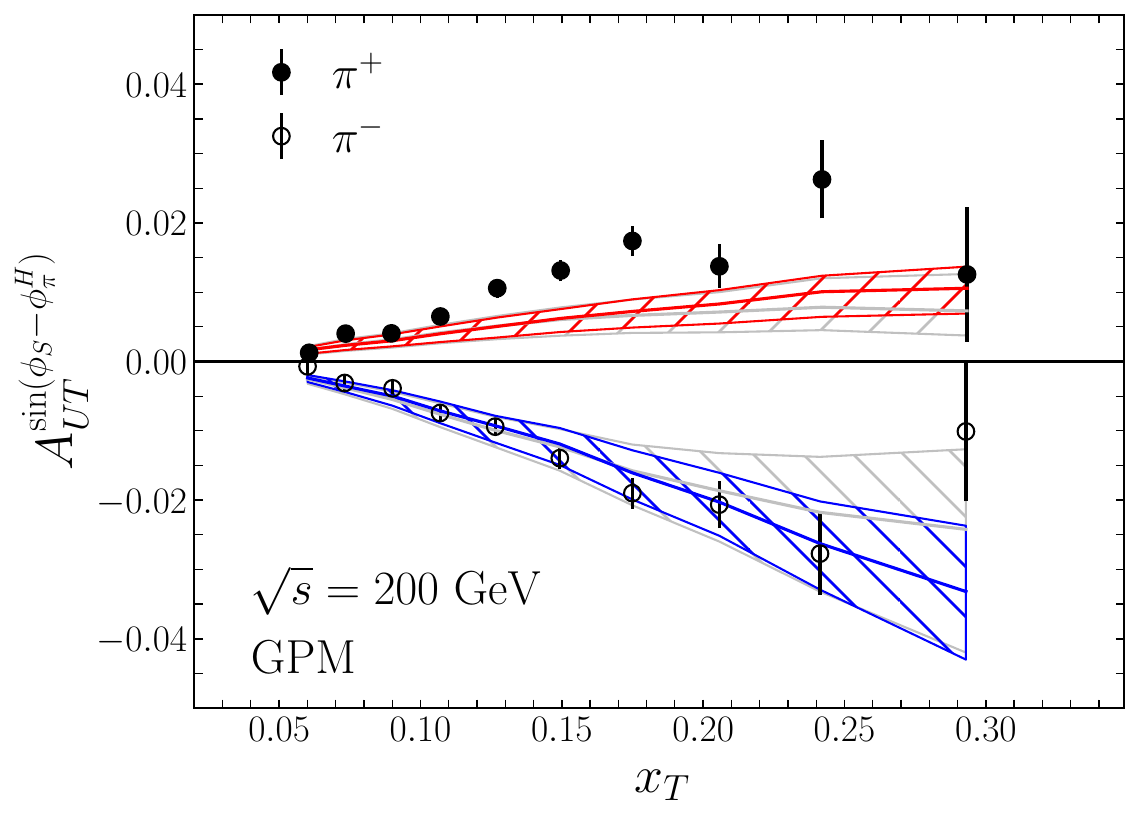}
\hspace*{-.1cm}
\includegraphics[width=4.35cm, keepaspectratio]{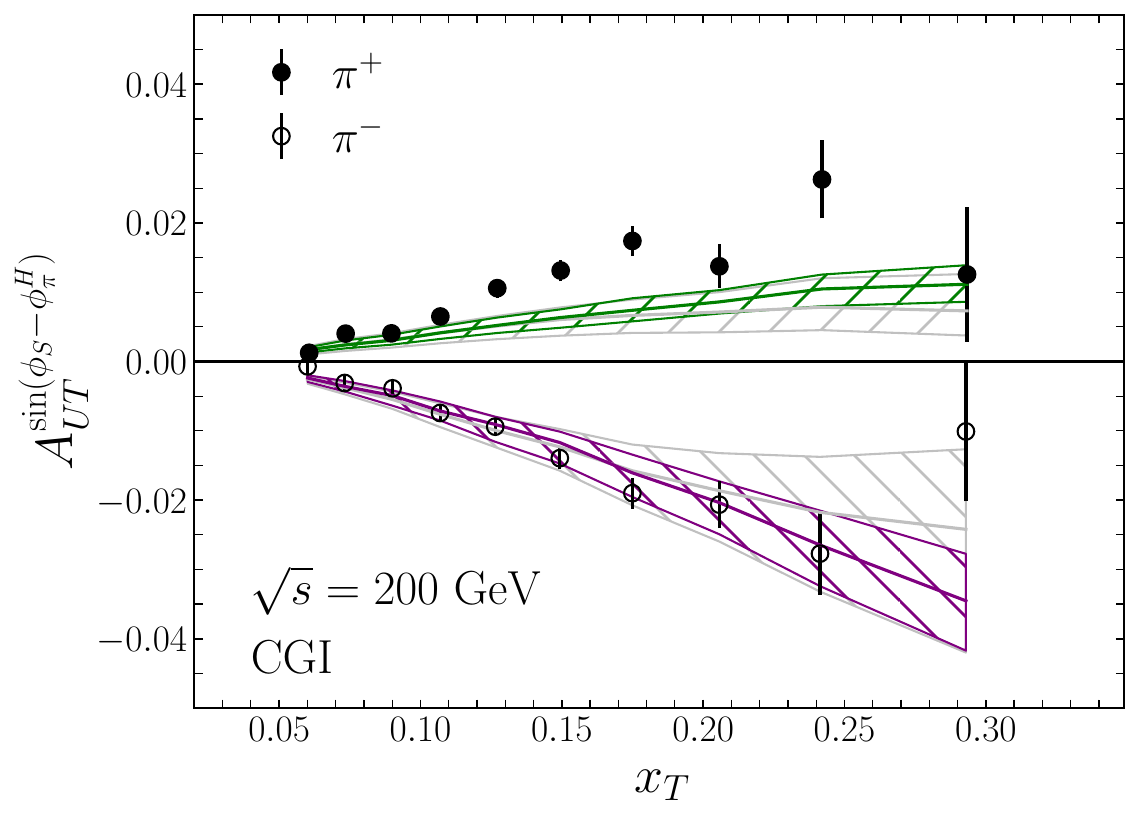}\\
\includegraphics[width=4.35cm, keepaspectratio]{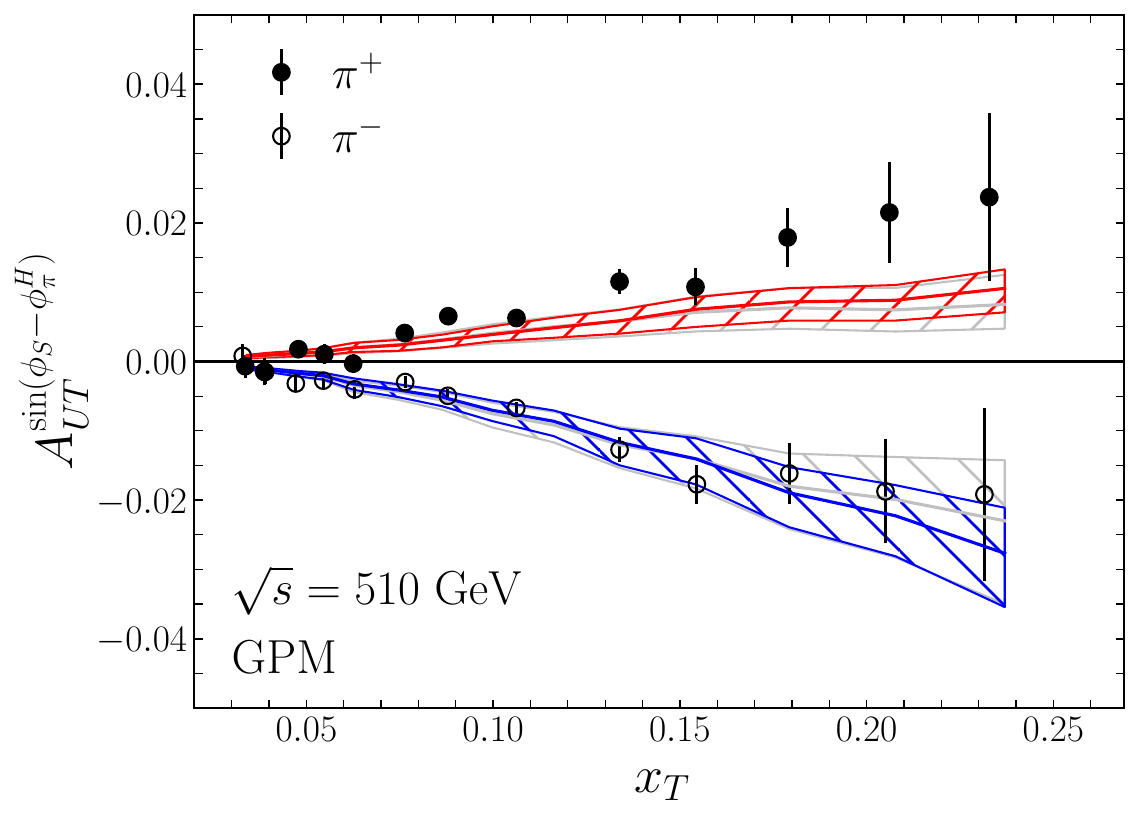}
\hspace*{-.1cm}
\includegraphics[width=4.35cm, keepaspectratio]{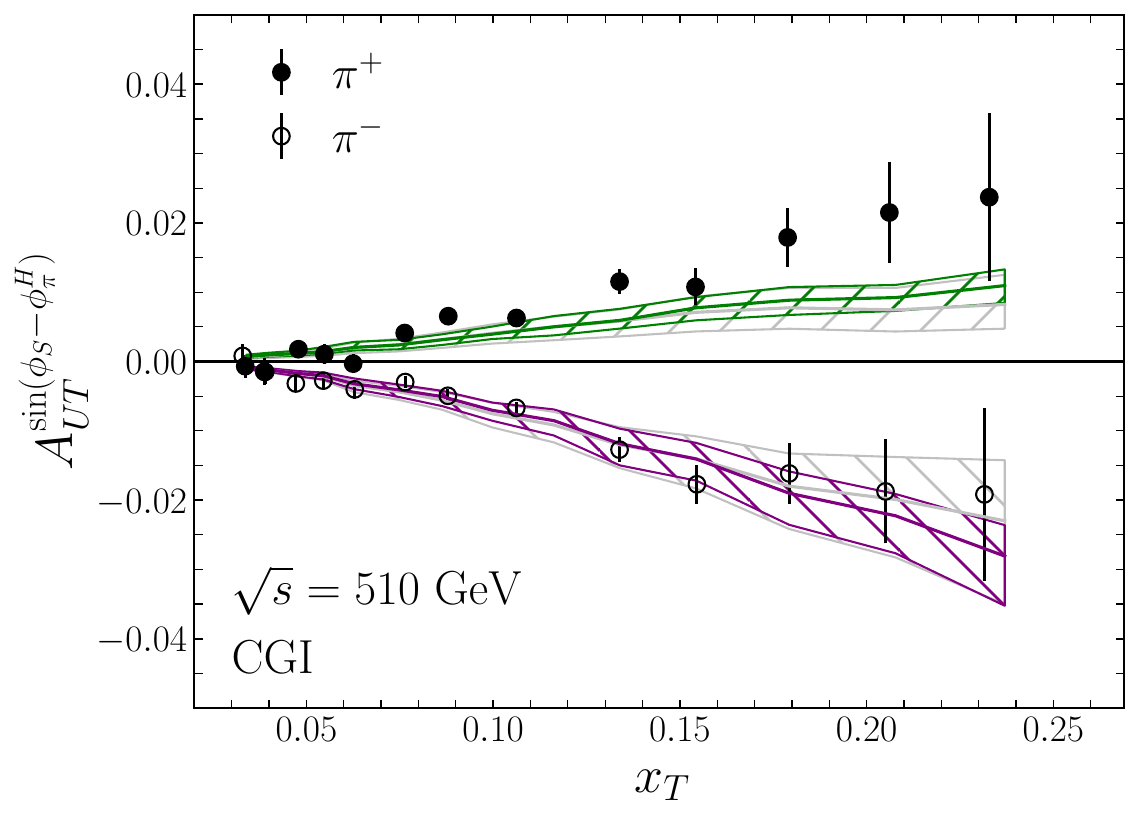}

\caption{Estimates of the Collins azimuthal asymmetry for $p^\uparrow p\to {\rm jet}\, \pi\, X$, based on the reweighted extractions of Ref.~\cite{Boglione:2024dal}, as a function of $x_T$ and integrated over $\eta_{\rm j}$ in the forward region, at $\sqrt s=200$~GeV  (upper panels) and at $\sqrt s= 510$~GeV  (lower panels) for the GPM (left panels) and CGI-GPM (right panels) formalisms. Uncertainty bands and data as in Fig.~\ref{fig:asymxt}.
}
\label{fig:asymxt-rew} 
\end{figure}

In Fig.~\ref{fig:asymz-rew} and Fig.~\ref{fig:asymxt-rew} we show our predictions for the Collins azimuthal asymmetries for the production of charged pions within a jet in $p^\uparrow p$ collisions, based on the reweighted extractions of Ref.~\cite{Boglione:2024dal}, as a function of $z$, at fixed $p_{{\rm  j} T}$, and as a function of $x_T$, integrated over $z$, respectively, at $\sqrt s = 200$ GeV upper panels) and $\sqrt s = 510$ GeV (lower panels) within the two approaches mentioned above (GPM on the left panels and CGI-GPM on the right panels). As one can see the agreement is still extremely satisfactory, even within the uncertainty bands, and, as expected, no significant differences appear for the two models. The same happens for all other cases, $j_T$-dependent asymmetries and the $\pjT$-binned ones (cfr.~Figs.~\ref{fig:asymzbins}-\ref{fig:asymjtbins}), not shown here for brevity. These findings, once again, support the possibility of a  unified TMD picture of SSAs for single-scale inclusive and two-scale less inclusive processes. 

\section{Conclusions and outlook}\label{sec:conclusions}

In this work we have shown how, by employing the transversity and the Collins functions as extracted from SIDIS and $e^+e^-$ annihilation processes, one is able to describe, in sign and size, the behavior of the available data from STAR Collaboration on the azimuthal asymmetries for pion-in-jet production in polarized $pp$ collisions. 
The overall agreement of our theoretical estimates when compared against two data sets, collected at different energies and spanning different $p_{{\rm j}T}$ values, is definitely remarkable. The almost energy independence of STAR data and predictions is another aspect that clearly emerges. 

With all due caution, coming from the assumptions and the simplified scheme adopted, this result supports the universality hypothesis of the Collins function and 
the validity of TMD factorization  for this kind of processes. In other words, no sizable signals of factorization breaking effects appear, at least when a collinear configuration for the initial state is assumed. At the same time, the impact of proper TMD evolution seems to play only a marginal role, definitely less relevant than the standard DGLAP collinear evolution. This behavior, as found in other phenomenological analyses and based on the experimental evidence, is somehow expected for azimuthal spin asymmetries, being ratios of cross sections.

The few small discrepancies between theoretical estimates and data, still appearing in this study, can be easily understood in terms of the simplified model adopted, together with the small number of parameters employed and the flavor-independent Gaussian $p_\perp$-functional form of the Collins function. These assumptions deserve further investigation and the model can be certainly refined by exploring more flexible parameterizations. Nevertheless, even with these words of caution, these findings are quite encouraging and represent another important step in our comprehension of azimuthal asymmetries in QCD.  

As a complementary investigation we have also considered an extraction of the transversity and the Collins FF where SSA data for inclusive processes are suitably included through a simultaneous reweighting procedure. Once again the agreement of the predictions and pion-jet data is good, indicating the reliability of a unified and global description of all these observables within a TMD approach. 
 
Further investigations are certainly necessary. From the theoretical point of view, a more accurate treatment of the jet fragmentation mechanism, beyond the LO accuracy, could improve the formulation of these observables in QCD, and help at the same time in shedding light on potential factorization-breaking effects. More generally, a global fit, or a reweighting procedure, including SIDIS and $e^+e^-$ azimuthal asymmetries, could help to better constrain the behavior of the Collins FF (at small and large $z$) as well as the transversity (at large $x$). Exploring more flexible parameterizations in the partonic variables, including the intrinsic transverse momentum dependence in the TMD-FFs, is another issue for future analyses. Similarly, the study of the Collins azimuthal asymmetries in pion-in-jet production in lepton-proton collisions, as those measurable at EIC, will be crucial to refine and extend the extraction of the transversity distribution also at low-$x$ values (probing the totally unknown sea region), test the overall picture  and provide a deeper comprehension of these phenomena. 

\section*{Acknowledgments}

We are very grateful to Qinghua Xu and Yixin Zhang for interesting discussions and for providing us with the STAR preliminary data and all details on the experimental cuts. We thank Francesco Murgia for his careful reading of the manuscript. The work of U.D.~and C.F.~is supported by the European Union ‘‘Next Generation EU’’ program through the Italian PRIN 2022 grant n. 20225ZHA7W. C.F.~is further supported by the European Union’s Horizon Europe research and innovation programme under the Marie Skłodowska-Curie grant agreement n.~101150792 (STAT-TMDs). The work of M.Z.~was partially supported by the U.S. Department of Energy contract No.~DE-AC05-06OR23177, under which Jefferson Science Associates, LLC operates Jefferson Lab, and conducted in part under the Laboratory Directed Research and Development Program at Thomas Jefferson National Accelerator Facility for the U.S. Department of Energy.

\bibliographystyle{elsarticle-num}
\bibliography{references}

\end{document}